\documentclass[9pt,twocolumn,twoside]{osajnl}
\journal{ol}
\setboolean{shortarticle}{true}
\graphicspath{{Figures/}}
\usepackage[justification=justified]{caption}

\title{Technique for Generating Broadband FM Light }

\author[1,*]{S. E. Harris}
\author[2]{Brandon Buscaino}

\affil[1]{Departments of Electrical Engineering and Applied Physics, Ginzton Laboratory, Stanford University, Stanford, California 94305, USA}
\affil[2]{Department of Electrical Engineering, Ginzton Laboratory, Stanford University, Stanford, California 94305, USA}

\affil[*]{Corresponding author: seharris@stanford.edu}

\begin{abstract}
We suggest a technique for using off-resonance spectral comb generation to produce broadband frequency modulated, and therefore, amplitude quieted light. Results include closed-form formulae for the amplitudes and phases of all of the spectral components.   
\end{abstract}

\setboolean{displaycopyright}{true}

\begin{document}

\maketitle

\section{Introduction}

Recent advances in integrated frequency comb technology, such as the demonstration of over 900 comb lines from dispersion-engineered, ultra-low loss thin-film LiNbO$_3$ resonators~\cite{Zhang} have regenerated interest in resonator-enhanced electro-optic devices. 
This Letter makes two contributions. The first is the description of a new method for generating a broad spectrum FM light signal. The second is the development of an analytical eigen-decomposition technique that allows a closed-form expression for the modal amplitudes and phases of a detuned spectral comb generator. Because the generated signal is frequency modulated, its amplitude fluctuations, as compared to a randomly phased signal with comparable bandwidth, are greatly reduced. 

 In 1964, on the same time scale as theoretical work by Gordon and Rigden \cite {Gordon} and Yariv \cite{Yariv}, workers attempting to use an electrooptic phase modulator to mode lock a He-Ne laser discovered that a phase modulator placed inside of the cavity and detuned from the modal resonances would produce a set of modes with Bessel function amplitudes, and called the resulting device an FM laser \cite{Harris-Targ}.  In 1972 resonator-enhanced electro-optic frequency combs were demonstrated by Kobayashi and colleagues~\cite{Kobayashi}. In the early 1990's work by Kourogi resulted in broadband comb generators \cite{Kourogi}. 
In 1993, application to fiber optic technology and a method for time-domain analysis were suggested by Ho and Kahn ~\cite{Ho}. And recently, Yuan and Fan have shown how a detuned-ring system might be used for unidirectional frequency translation \cite{Yuan}. 

\section{Input-Output Relations}

\begin{figure}[t!]
\centering
\includegraphics[width=\linewidth]{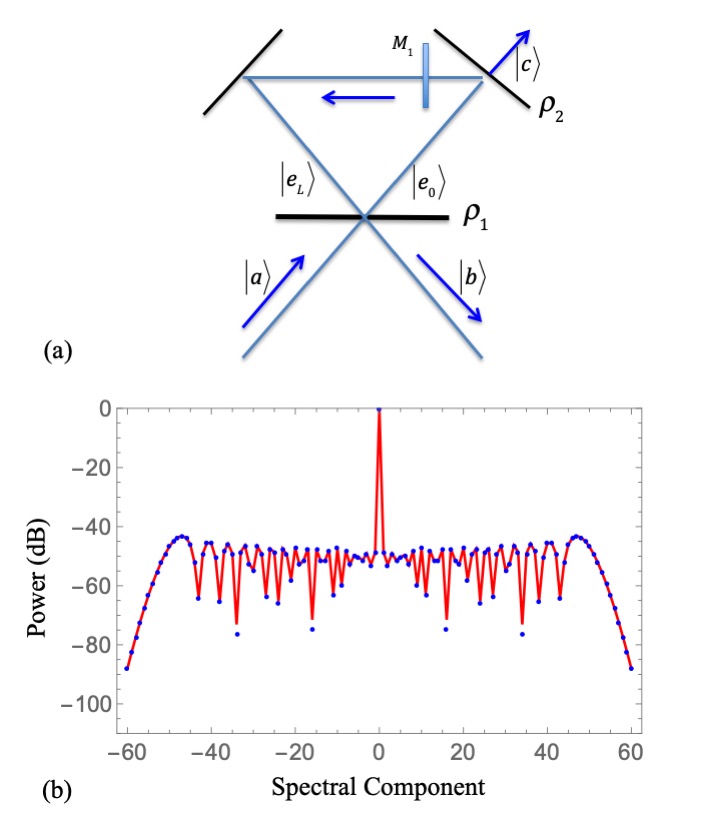}
\caption{(a) Schematic of a phase modulator, $M_1$, embedded in a ring cavity; (b) Output spectrum from port $B$.  For appropriate conditions the light observed from port $C$ will approximate a broadband FM signal. The red curve is obtained by solution of 241 linear equations, Eqs.~(\ref{eq1}-\ref{eq2}). The blue dots are the approximate solution, Eqs.~(\ref{components}).  The width of the spectrum is 25 times larger than that produced by a phase modulator without a resonator. Parameters: $\delta=2$, $\Gamma=50$, $\xi=\rho_1=\rho_2=\sqrt{0.999}$, and $\beta''=0$.}
\label{Fig1}
\end{figure}

The system considered here is shown in Fig.~\ref{Fig1}(a), where a monochromatic plane wave is incident onto a ring resonator with an electro-optic phase modulator near the right hand turning mirror. A numerical example of the generated spectrum is shown by the red curve in Fig. \ref{Fig1}(b). We derive a closed-form expression, Eq. (\ref{components}), that approximates the numerical spectrum, and is shown by blue dots in Fig. \ref{Fig1}(b).

We work in the frequency domain and express all fields as Dirac kets. For example the reflected field $|b\rangle$ has spectral components $b_n=\langle n|b\rangle$ with frequencies 
$\omega_0+n \omega_m$ where $\omega_m$ is the frequency driving the electro-optic modulator $M_1$ and  $\omega_0$ is the  incident optical frequency which is exactly on resonance with a particular optical mode. We take these spectral components to extend from $-N_m$ to  $+N_m$, with a total of $2 N_{m}+1 $ spectral components. The temporal quantity associated with the ket $ |b\rangle$ is $b(t)=\sum_{n=-N_m}^{N_m} b_{n} \exp[i n \omega_{m} t]$.

We write input and output boundary conditions at the beam splitter~\cite{Raymer}. The quantities $\rho_1$ and $\rho_2$  are the electric field reflection coefficients at the the beam splitter and right hand turning mirror. $\tau_1$ and $\tau_2$ are the corresponding transmission coefficients and satisfy $\rho_{i}^2+\tau_{i}^2=1$, with a convention of a $\pi$ phase shift on reflection from outside of the splitter. To account for loss, we define $\xi =\exp(-\alpha L)$, where $\alpha$ is the per-length electric field absorption coefficient. When $\xi = 1$ there is no loss, and when $\xi=0$, there is no transmission.  

The input-output conditions may then be written as:

\begin{equation}
\begin{aligned}
&|e_0\rangle= \tau _1 |a\rangle +\rho _1 |e_L\rangle\\
&|e_L\rangle=\xi  \rho _2 M_2 M_1 |e_0\rangle\\
&|b\rangle= \tau _1 |e_L\rangle -\rho _1 |a\rangle\\
&|c\rangle= \tau _2 |e_0\rangle
\label{eq1}
\end{aligned}
\end{equation}

\noindent
where the location $0$ is shorthand for $0+$, i.e. inside the ring immediately following the beam splitter. The location L is shorthand for $L_{-}$, i.e. just before the beam splitter. The ket $|a\rangle$ represents the input driving field and in this work will be taken as a monochromatic field with unit amplitude at a center driving frequency $\omega_0$ so that $\langle n|a\rangle=\delta_{n,0}$.

A sinusoidal phase modulator with peak phase retardation $\delta$ multiplies each frequency component at its input by $\exp [i\delta \cos (\omega_{m} t)]= \sum_{q=-\infty}^{+\infty} i^{q} J_{q} (\delta) \exp[i q\omega_{m} t]$. This modulator is described by the operator $M_1$ with matrix elements  $\langle n| M_1 |r\rangle =i^{n-r} J_{n-r}(\delta)$.  Denoting an on-resonance center frequency as $\omega_0$, and  with  $T=L/V_{g} (\omega_0) $ as the transit time around the ring, the free spectral range of the ring resonator is $\omega_{fsr}=2\pi/T$. We take the modulation frequency $\omega_m= \omega_{fsr}+\delta\omega$ with $|\delta \omega|<\omega_{fsr}$. With $\beta''$ as the dispersive parameter, the collected phase through the ring is described by an operator $M_2$ with diagonal matrix elements $\langle n| M_2 |r\rangle =\exp \left [-i (n \delta \omega  T+ \omega_{m}^2  n^2   \beta'' L/2)\right] \delta_{n,r}$. The operators $M_1$ and $M_2$ are unitary and do not commute. We define the matrix $M=\xi \rho_1 \rho_2 M_{2} M_{1}$ with components $\langle n| M  |r\rangle =(\xi \rho_1 \rho_2)  i^{n-r} e^{-i n \delta \omega T} J_{n-r}(\delta )$. Eqs.~(\ref{eq1}) then combine to

\begin{equation}
\begin{aligned}
&|e_ {L} \rangle = \frac{\tau_1}{ \rho_1} (I-M)^{-1} M  |a\rangle \\
&|e_ {0} \rangle =\tau_1 (I-M)^{-1} |a\rangle
\label{eq2}
\end{aligned}
\end{equation}

Eqs. (\ref{eq1}-\ref{eq2}) are not restricted to high-{\it{Q}} cavities and are valid for all values of the mirror reflectivities and loss. In the absence of loss ($\xi=1$), total power is conserved. Similar equations have been previously derived~\cite{Kovacich, Buscaino}. Notably, if the loss is zero and the mirror reflectivities are unity, one eigenvalue of $(I-M)$ is zero and $(I-M)^{-1}$ does not exist. With loss and non-zero reflectivity included, the smallest eigenvalue of $(I-M)$ is $\lambda_0=(1-\xi \rho_1 \rho_2)$ (see Supplementary Material). 

\section{FM Laser}

We show that the detuned comb generator has much in common with the FM laser~\cite{Harris-McDuff}. To obtain a linearized solution for the FM laser we neglect saturation, set the loss to zero, and the cavity reflectivity to unity.   
We assume that $\delta \omega T$ is sufficiently small such that $\exp[-i n \delta \omega T] \simeq (1-i n \delta \omega T)$. This assumption assures that the instantaneous single pass frequency shift through the modulator is equal to the frequency shift of the broadband FM signal with index $\Gamma$~\cite{Kuizenga}. For small $\delta$ we obtain

\begin{equation}
\begin{aligned}
&(n T \delta \omega) E_{n}=\frac{\delta}{2} (E_{n-1}+E_{n+1})\\
&E_{n}= J_{n}(\Gamma); \ \  \Gamma=\frac{\delta}{\delta \omega T}
\label{eq3}
\end{aligned}
\end{equation}

In the time domain, the Fourier components $E_{n}$ correspond to a frequency-modulated wave $\cos[\omega_{0} t+\Gamma \cos (\omega_{m} t)]$ with instantaneous frequency $\omega_{0} -\Gamma \omega_m\sin(\omega_m t)$. The spectral bandwidth is approximately $2 \Gamma \omega_m$ and the number of spectral components with appreciable amplitude is $2 \Gamma$. In general the temporal dynamics of the FM laser are much more complex. Especially when the gain profile is flat, there are many FM signals centered on different cold cavity modes that compete with each other~\cite{Harris-McDuff, Kuizenga}.

\section{Approximate Solution of Equations (1-2)}
 
With $ |q \rangle$ and $\lambda_q$ as the orthonormal eigenvectors and eigenvalues of the operator $(I-M)$, we can express its inverse as

\begin{equation}
\left (I-M \right )^{-1}=\sum_{q} \frac{1}{\lambda_q} | q \rangle \langle q | 
\label{eq4}
\end{equation}

\noindent
If one eigenvalue, $\lambda_0$, is much smaller than the others, then

\begin{equation}
\left (I-M \right )^{-1}\cong \frac{1}{\lambda_0} | q_0 \rangle \langle q_0 | 
\label{eq5}
\end{equation}

\noindent
We now assume that the eigenvector $ | q_0 \rangle $ is the same as that of the FM laser with an additional time delay factor dependent on the position of the phase modulator in the ring. The corresponding eigenvalue 
 $\lambda_0=(1-\xi\rho_1\rho_2 )$ is such that, when $\delta \omega T<0.1$,

\begin{equation}
\begin{aligned}
&\left( I-M\right) | q_0 \rangle=\lambda_0 | q_0 \rangle \\
& \langle n|q_0\rangle=J_n(\Gamma) \exp{(-i n  \delta \omega T /2)}
\label{eq6}
\end{aligned}
\end{equation}

\noindent
From the first of Eqs. (\ref{eq2}) we have
\begin{equation}
|e_L\rangle=\left(\frac{\xi \tau_1 \rho_2 }{\lambda_0 }\right) \left<q_0| M_2 M_1 | a \right>  |q_0 \rangle 
\label{eq7}
\end{equation}

\noindent
Using Graf's sum rule, we have shown that 

\begin{equation}
\begin{aligned}
\langle q_{0}| M_2 M_1 |a\rangle &=\sum _{q}  i^{-q} \exp (i q \delta \omega T/2) J_q(\Gamma ) J_q(\delta )  \\
&=J_{0} \left(Z\right) 
\label{eq8}
\end{aligned}
\end{equation}

\noindent
where, noting $\delta \omega T = \delta / \Gamma$,

\begin{equation}
Z=\sqrt{(\Gamma^2+\delta^2-2\Gamma \delta \sin(\delta / (2 \Gamma))} 
\label{eq9}
\end{equation}
When $\delta<<\Gamma$, as in most of this work, $Z\rightarrow\Gamma$. Combining Eqs. (\ref{eq7}-\ref{eq9}), the approximate solution to Eqs. (\ref{eq1}-\ref{eq2}) is 

\begin{equation}
\begin{aligned}
&|e_0\rangle=\tau_1 |a \rangle +\left(\frac{\xi \tau_1 \rho_1 \rho_2 }{\lambda_0 }\right) J_{0}(Z)  |q_0 \rangle \\
&|e_L\rangle=\left(\frac{\xi \tau_1 \rho_2 }{\lambda_0 }\right) J_{0}(Z)  |q_0 \rangle \\
&|b\rangle= \tau _1 |e_L\rangle -\rho _1 |a\rangle\\
&|c\rangle= \tau _2 |e_0\rangle
\label{eq10}
\end{aligned}
\end{equation}

\begin{figure}[tbp]
\begin{center}
\includegraphics[width=3.45in]{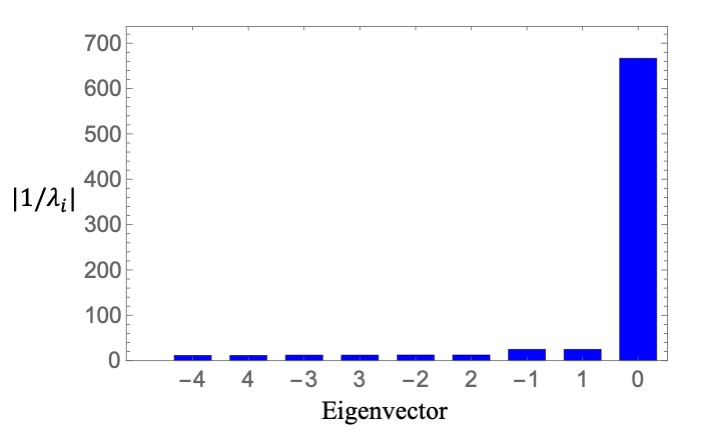}
\end{center}
\caption {Inverse absolute value of the nine smallest eigenvalues of  $(I-M)$. The parameters are the same as those of Fig.~\ref{Fig1}.}
\label{Fig2}
\end{figure}

\noindent
Using Eq. (\ref{eq7}), the field components are then
\begin{equation}
\begin{aligned}
&b_n=-\rho_1\delta_{n,0}+ \frac{\xi \tau_{1}^2 \rho_2}{\lambda_{0}} J_{0}(Z) J_{n}(\Gamma) \exp \left(-i n \delta \omega T/2\right) \\
&c_n=\tau_1 \tau_2 \delta_{n,0}+ \frac{\xi \tau_{1}  \tau_{2} \rho_1  \rho_2}{\lambda_{0}} J_{0}(Z) J_{n}(\Gamma) \exp \left(-i n \delta \omega T/2\right) 
\label{components}
\end{aligned}
\end{equation}

\noindent
With unity input power, the powers from port B and port C are

\begin{equation}
\begin{aligned}
&P_{B}=\rho _1^2+\frac{\xi  \rho _2 \tau _1^2 J_0(Z) \left(\xi  \rho _2 \tau _1^2 J_0(Z)-2 \lambda _0 \rho _1 J_0(\Gamma )\right)}{\lambda _0^2}  \\
&P_{C}=\frac{\tau _1^2 \tau _2^2 \left(\lambda _0^2+\xi  \rho _1 \rho _2 J_0(Z) \left(2 \lambda _0 J_0(\Gamma )+\xi  \rho _1 \rho _2 J_0(Z)\right)\right)}{\lambda _0^2}
\label{power}
\end{aligned}
\end{equation}

\begin{figure}[t]
\begin{center}
\includegraphics[width=3.45in]{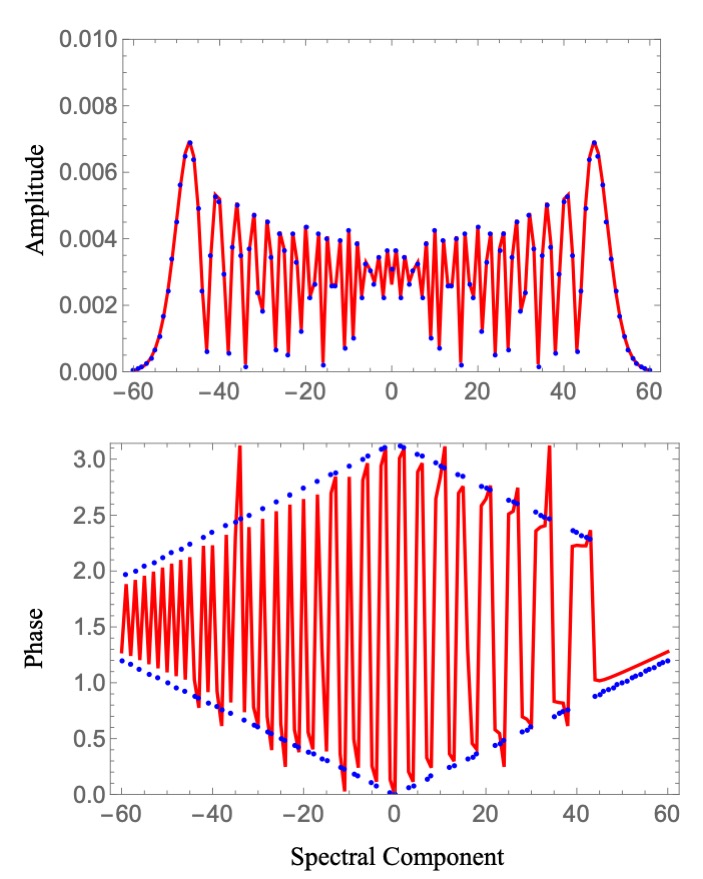}
\end{center}
\caption{Amplitude and absolute value of the phase of spectral components at port C. The red curve is the result of the solution of 241 linear equations, Eqs. (\ref{eq1}-\ref{eq2}). The blue dots are the approximate solution, Eq. (\ref{components}). Parameters are identical to Fig.~\ref{Fig1}.}
\label{Fig3}
\end{figure}

\section{Criterion for validity}

When $q\neq 0$, the reflectivities are unity, and the loss is zero, the eigenvalues $\lambda_q$ are  $q \delta \omega T$ (see Supplementary Material). Therefore, a rough criterion for the validity of the approximate solution of Eqs.~(\ref{components}) is that $\lambda_{0}$ must be small as compared to  $\lambda_{q} $, i.e.

\begin{equation}
\eta=\frac{(1- \xi  \rho_1 \rho_2)}{\delta \omega T} << 1
\label{eq13}
\end{equation}

\noindent
We may also understand this criterion by noting that the enhancement in the modulation index, $\Gamma/\delta=1/(\delta \omega T)$, must be less than the number of round trip passes allowed by the combined loss and  reflectivity, $1/ (1-\xi \rho_1 \rho_2 )$.
This criterion determines the region of validity for the approximate solution and, of more importance, the parameters necessary for generating FM light.

\section{Examples and consequences}

Fig.~\ref{Fig2} shows the inverse of the nine smallest eigenvalues of the matrix $(I-M)$. In this figure, the parameters are the same as in Fig.~\ref{Fig1}. As expected,  $\lambda_0=(1-\rho_1 \rho_2 \xi)$ exactly, and the ratio of  $1/ \lambda_0$ to $1/ \lambda_{\pm 1}$ is $26.6733$, as compared to the predicted value $(\delta \omega T)/ (1-\rho_1 \rho_2 \xi)=26.6703$ (see Supplementary Material).                           

Fig.~\ref{Fig3} shows the Fourier amplitudes and phases of the output field at port C. The numerical solution is shown in red and calculated from Eqs. (\ref{eq1}-\ref{eq2}). The approximate solution is shown as blue dots and calculated from Eqs. (\ref{components}). Because $\eta=.037$ is much less than unity, the agreement between the exact solution and Eqs. (\ref{components}) is good.  In general we find that if one is only interested in the amplitude, and not the phase, then significantly higher values of $\eta$ still yield good agreement. 

\begin{figure}[t]
\begin{center}
\includegraphics[width=3.45in]{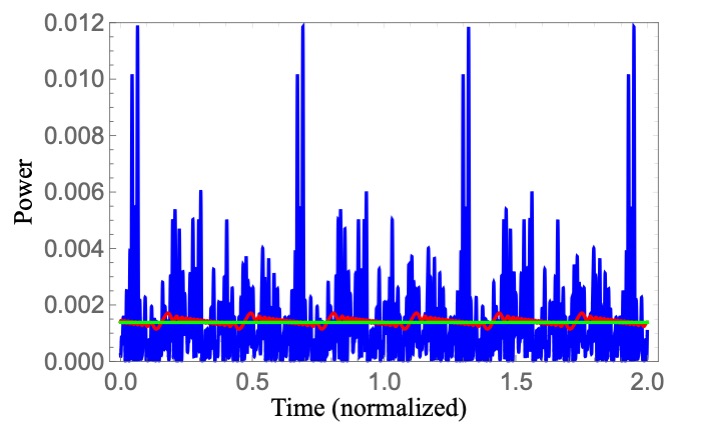}
\end{center}
\caption{Fourier synthesis of the temporal output at port C.  The blue curve is obtained by assigning random phases to each of the modal amplitudes from Eqs. (\ref{eq1}-\ref{eq2}). These amplitudes and phases are Fourier-synthesized to form a temporal profile. The absolute value and square of this profile is plotted here. The red curve is obtained by retaining the phases, as well as the amplitudes, of the numerical solution of Eqs. (\ref{eq1}-\ref{eq2}). The green (flat) line is the result of the approximate solution of Eq. (\ref{components}) when $\eta \rightarrow 0$. Other parameters are the same as in Fig. \ref{Fig1}. }
\label{Fig4}
\end{figure}

In Fig. \ref{Fig4}, we sum the frequency domain components $c_n=\langle n|c \rangle$ to obtain $c(t)$ and plot $|c(t)|^2$ as a function of time (red curve). For comparison we plot the square of the absolute value of the time domain function that results if the same modal amplitudes have random phases; i.e. the phases are randomly assigned to determine the temporal function which is then squared and plotted (blue curve). Here, as in Figs.~\ref{Fig1}-\ref{Fig3}, the parameter $\eta=0.037$ and the reduction in intensity variation, as compared to the randomly phased function is substantial. If the loss and reflectivity is reduced further, $\eta \rightarrow 0$, and the temporal profile becomes flat (green curve), as obtained from Eqs. (\ref{components}).
\noindent

\begin{figure}[tbp]
\begin{center}
\includegraphics[width=3.45in]{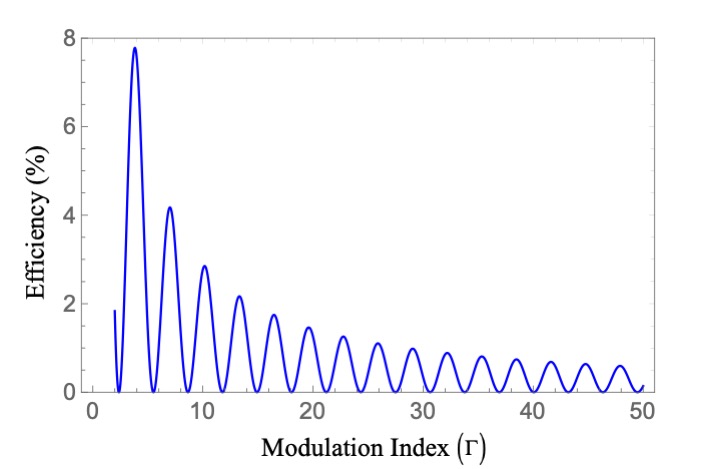}
\end{center}
\caption{Power conversion efficiency at port C as obtained from Eqs. (\ref{power}). With $\delta \omega T=\delta/ \Gamma$, the efficiency is independent of $\delta$. Other parameters are the same as in Fig. \ref{Fig1}.}
\label{Fig5}
\end{figure}
Though not developed here, one may also show that the instantaneous phase $\phi(t)$, summed over all spectral components, is  
\begin{equation}
\phi(t)=\Gamma \sin \left(\omega_m t-\frac{\delta \omega T }{2}\right)
\end{equation}
This phase is consistent with a total spectral bandwidth of $2 \omega_m \Gamma$.

In Fig.~\ref{Fig5} we use Eqs. (\ref{power}) to plot the efficiency, i.e. the ratio of the power from port C to the input power. In this plot, the modulator drive strength $\delta$ is fixed at 2 radians and $\delta \omega$ is varied so that $\Gamma$ varies over the range $2\leq\Gamma\leq50$. We find, somewhat surprisingly, that this same efficiency curve is obtained for arbitrary $\delta$. For example, the predicted efficiency peaks at about $7.78\%$ when $\Gamma= 3.83$. For this $\Gamma$, at $\delta=0.2$ we find an efficiency of $7.78\%$, and for a $\delta=2$ an efficiency of $7.76\%$. This excellent agreement is the result of the assumption of a small power loss and a reflectivity of $0.1\%$.
\section{Summary}
This work suggests a technique for generating a frequency-modulated, and therefore, an amplitude-quieted broadband waveform. Such a light source may be used to seed an amplifier to avoid the nonlinearities or damage that would otherwise be caused by a mode-locked or randomly phased source. 
As compared to a monochromatic input, an FM generator might also be used to increase the power which may be extracted from a broadband inhomogeneous amplifying medium. Closed-form formulae have been shown to be in good agreement with the result of the numerical solution of hundreds of simultaneous equations.

\section{Supplemental Material}
Here, we make sufficient approximations to derive analytical expressions for the eigenvalues and eigenvectors of the matrix M. These quantities provide guidance for the key assumption, Eq. (\ref{eq5}), and also for the criterion of Eq. (\ref{eq13}) .

Starting with the matrix $(I-M)$ we expand it in a power series to first order in $\delta$ and $\delta \omega$, and let $\delta\times(\delta \omega)=0$. The result is a reduced matrix $M_R$ with non-zero matrix elements
\begin{equation}
\begin{aligned}
& \left <r | M_{R} | r \right > =(1-\rho_1 \rho_2  \xi)+i ( r  \rho_1 \rho_2  \xi \delta \omega T)  \\
&\left <r | M_{R} | r_{\pm 1} \right > =-i \rho_1 \rho_2  \xi  \frac{\delta}{2}
\end{aligned}
\end{equation}
With the reflectivity set to unity and no loss $(\rho_1=\rho_2=\xi=1)$, the eigenvalues $\lambda_{q} $ and the eigenvectors $|q\rangle$ of $M_R$ are
\begin{equation}
\begin{aligned}
&\lambda_{q} =i q  \delta \omega T \\
&\langle n|q \rangle=J_{n-q}( \Gamma)
\end{aligned}
\end{equation}
where $\Gamma=\delta/ (\delta \omega T)$. At this point, $\lambda_{0}=0$ and the solution for the FM laser, Eq. (\ref{eq3}), is the eigenvector $|q_0\rangle$.

To incorporate non-zero loss and finite reflectivity we use first order perturbation theory; i.e., define a perturbative matrix 
\begin{equation}
\Delta M_{R}=M_{R}-M_{R}|_{\rho_1 \rho_2  \xi=1}
\end{equation}
so that 
\begin{equation}
\lambda_{0}= \Delta \lambda_{0}= \langle q_0|\Delta M_{R} |q_0 \rangle=(1-\rho_1 \rho_2  \xi)
\end{equation}
Of note, using $\sum_{n=-\infty} ^ {\infty} J_{n-r} (\zeta) J_{n-p} (\zeta)=\delta_{r,p}$, the eigenvectors $ |q\rangle$ form a complete orthonormal set. The result $\lambda_0=(1-\rho_1 \rho_2 \xi)$ holds  true for the matrix $M$ as well as for $M_R$ and does not require perturbation.

\section*{Funding Information}

This work was supported by: Facebook, Inc., Maxim Integrated, and the National Science Foundation (Grant ECCS-1740291 E2CDA).

\section*{Disclosures}

The authors declare no conflicts of interest.

\section*{Acknowledgments}

The authors acknowledge helpful discussions with Joseph M. Kahn, Carsten Langrock, Marty Fejer, Shanhui Fan, and David Miller. 

%
%




\end{document}